# Deep learning for the rare-event rational design of 3D printed multi-material mechanical metamaterials

H. Pahlavani[a,1], M. Amani[a], M. Cruz Saldívar[a], J. Zhou[a], M. J. Mirzaali[a], A. A. Zadpoor[a]

[a] *Department of Biomechanical Engineering, Faculty of Mechanical, Maritime, and Materials Engineering, Delft University of Technology (TU Delft), Mekelweg 2, 2628 CD, Delft, The Netherlands*

---

[1] Corresponding author. Tel.: +31-613221032, *e-mail:* h.pahlavani@tudelft.nl

**ABSTRACT**

Emerging multi-material 3D printing techniques have paved the way for the rational design of metamaterials with not only complex geometries but also arbitrary distributions of multiple materials within those geometries. Varying the spatial distribution of multiple materials gives rise to many interesting and potentially unique combinations of anisotropic elastic properties. While the availability of a design approach to cover a large portion of all possible combinations of elastic properties is interesting in itself, it is even more important to find the extremely rare designs that lead to highly unusual combinations of material properties (*e.g.*, double-auxeticity and high elastic moduli). Here, we used a random distribution of a hard phase and a soft phase within a regular lattice to study the relationship between the random distribution of both phases and the resulting anisotropic mechanical properties of the network in general and the abovementioned rare designs in particular. The primary challenge to take up concerns the huge number of design parameters and the extreme rarity of such designs. We, therefore, used computational models and deep learning algorithms to create a mapping from the space of design parameters to the space of mechanical properties, thereby (i) reducing the computational time required for evaluating each design to $\approx 2.4 \times 10^{-6}$ s and (ii) making the process of evaluating the different designs highly parallelizable. Furthermore, we selected ten designs to be fabricated using polyjet multi-material 3D printing techniques, mechanically tested them, and characterized their behavior using digital image correlation (DIC, 3 designs) to validate the accuracy of our computational models. The results of our simulations show that deep learning-based algorithms can accurately predict the mechanical properties of the different designs, which match the various deformation mechanisms observed in the experiments. The presented approach enables the evaluation of $\approx 10^{12}$ designs per month with a mid-range workstation, paving the way for the discovery of multi-material mechanical metamaterials with very rare combinations of anisotropic elastic properties.

# 1. INTRODUCTION

The rational design of architected materials with anisotropic properties enables them to offer optimal, multi-functional performance. For example, nature uses evolutionarily optimized micro-architectures to combine extremely high stiffness (in selected directions) with a light weight (*e.g.*, in wood and bone [1], [2], [3]) or to combine ultrahigh stiffness values with ultrahigh toughness (*e.g.*, in nacre [4], [5], [6]). In man-made designer materials that are also known as metamaterials, other combinations of mechanical properties may be sought, as they allow for devising novel functionalities. For example, a combination of auxetic behavior in various orthogonal directions and high stiffness is instrumental for the structural applications of auxetic materials [7], [8], [9].

To achieve a desired combination of material properties, the primary challenge is to find the specific micro-architectures that give rise to the desired properties. Once the micro-architecture is determined, the metamaterial can be fabricated using additive manufacturing (=3D printing) techniques. The recent emergence of powerful multi-material 3D printing techniques means that the micro-architecture not only consists of rationally designed, complex geometries but can also combine multiple materials with different mechanical properties. Many other design features found in nature, such as hierarchical micro-architectures [10], [11], [12], [13], functional gradients (in terms of both geometries and material properties) [14], [15], and soft-hard composites (similar to the organic and mineral phases in bone [11], [16], [17]) can be also realized to expand the range of the achievable properties.

Given such a wide range of possibilities for the fabrication of metamaterials with complex (multi-scale) geometries and complex spatial distributions of material properties, the space of possible design parameters is formidably large. Optimizing the design parameters is, therefore, challenging and requires an excessively large number of computational models to be solved. Such simulations are required not only to understand how the design parameters relate to the anisotropic elastic properties but, more importantly, to discover the very rare designs that give rise to the desired properties. For example, double-auxeticity (*i.e.*, auxetic properties in two orthogonal directions) is very rare (*i.e.*, as little as < 1.6% of the possible designs) in two-dimensional lattices [18]. Combining the double-auxeticity with the additional requirement of possessing high stiffness values in (both) directions results in the excessive rarity of micro-architectures that satisfy the design requirements.

Computational models, therefore, need to scan a vast design space to find rare events. Due to the "curse of dimensionality" [19], the number of designs that need to be evaluated is so large (*e.g.*, $10^{10}$-$10^{20}$) that extremely fast models and highly parallelizable algorithms are required. Computational models, such as finite element models, are not fast enough for that purpose. Here, we used deep learning to establish a mapping from the space of design parameters to that of the anisotropic elastic properties, thereby decreasing the solution time to $\approx 2.4 \times 10^{-6}$ s while also making the evaluation process extremely parallelizable. Recent progress in machine learning has led to significant achievements in different scientific fields [20], including the design of composites and metamaterials [21], [22], [23], [24], [25], prediction of material properties [26], [27], and optimization of manufacturing processes [28], [29]. However, the benefits from machine learning have not yet been demonstrated in the case of designing multi-materials mechanical metamaterials to achieve target properties.

The present research aimed to use computational models and deep learning models to predict the mechanical properties of multi-material mechanical metamaterials. We used planar lattices based on the re-entrant, cubic, and honeycomb unit cells (corresponding to the cell angles of 60°, 90°, and 120°, respectively) with random distributions of hard and soft phases. The ratio of the hard phase volume to the soft phase volume was varied as well. Computational (*i.e.,* finite element) models were then created to generate the training dataset required for the training of a deep learning model. Moreover, we selected three designs (one from each unit cell angle of 60°, 90°, and 120°) to be fabricated using an advanced multi-material 3D printing technique and applied digital image correlation (DIC) to measure the full-field strain patterns during the mechanical testing of the fabricated specimens. After training, the deep learning models were used to predict the elastic properties of a wide range of lattices ($1.5 \times 10^9$ different designs), given their design parameters. We also studied various combination of tiled designs (*e.g.*, four-tile and nine-tile structures) to show how combining multiple instances of these random lattices into a hybrid, tiled lattice can boost the possible range of mechanical properties. A fabrication and mechanical testing procedure similar to the one mentioned above (but without DIC) was applied to experimentally characterize seven additional tiled designs (*i.e.,* four four-tile structures and three nine-tile structures).

## 2. Materials and methods

We considered planar lattices with three groups of unit cell angles representing the negative (re-entrant, $\theta = 60°$), zero (orthogonal, $\theta = 90°$), and positive (honeycomb, $\theta = 120°$) values of the Poisson's ratio (Fig. 1a). We kept the overall dimensions of each design ($W, C$) as well as the dimensions of the constituent unit cells ($w, c$) unchanged. All three groups of designs were composed of 5×5 unit cells with similar in-plane ($t$) and out-of-plane ($T$) thicknesses. The geometrical parameters of the designed lattice structures are presented in Table S1 of the supplementary document. The hard and soft phases were randomly assigned to the struts of the structure so as to achieve various ratios of the volume of the hard phase to that of the soft phase ($\rho_h$ (%) = 5, 10, 20, 30, 40, 50, 60, 70, 80, 90, and 95). To further expand the space of possible mechanical properties, we studied the various combination of the unit cells. Moreover, we studied the stress distribution within the soft and hard elements of the unit cells and used a more uniform distribution of stresses as the criterion for selecting the best designs among all the designs with similar elastic properties.

### 2.1. Computational models

All finite element models were created using Matlab (Matlab R2018b, Mathworks, USA) codes. The codes were used to design the three groups of lattice structures (composed of unit cells with the three different cell angles of 60°, 90°, and 120°), to randomly assign the hard and soft phases to the struts of each design, and to perform the finite element simulations that estimate their mechanical properties (*i.e.,* elastic modulus and Poisson's ratio in two orthogonal directions). Our codes were further extended to combine single unit cell designs into four-tile and nine-tile lattice structures. In each structure, the adjacent designs were connected using a row of struts made of the hard material.

We used three-node quadratic beam elements (Timoshenko beam elements) with rectangular cross-sections and with two translational (*i.e.*, $u_x$, $u_y$) and one rotational (*i.e.*, $u_z$) degrees of freedom (DOF) at each node. We assigned elastic materials to both soft and hard phases with a similar Poisson's ratio of 0.48 but vastly different Young's moduli of 0.6 and 60 MPa (*i.e.*, $\frac{E_h}{E_s} = 100$), respectively. To estimate the mechanical properties of each structure in both the x- and y-directions, a strain of 3% in each direction was separately applied to the structure. Towards this aim, in one model, the top nodes were subjected to a strain of 3% in the y-direction ($u_x = u_z = 0$ and $u_y = 3\%$ strain), while all the degrees of freedom of the bottom nodes were constrained ($u_x = u_y = u_z = 0$). In the other model, the right nodes were subjected to 3% strain in the x-direction ($u_y = u_z = 0, and\ u_x = 3\%$ strain), while all the degrees of freedom of the left nodes were constrained ($u_x = u_y = u_z = 0$). The element stiffness matrix transferred to the global coordinate ($K^e$) was calculated as [30]:

$$K^e = Q^T \bar{K}^e Q, \qquad (1)$$

$$\bar{K}^e = \frac{E}{(1+\mu)} \begin{bmatrix} A(1+\mu)/L & 0 & 0 & -A(1+\mu)/L & 0 & 0 \\ 0 & 12I/L^3 & 6I/L^2 & 0 & -12I/L^3 & 6I/L^2 \\ 0 & 6I/L^2 & 4I(1+\mu/4)/L & 0 & -6I/L^2 & 2I(1-\mu/2)/L \\ -A(1+\mu)/L & 0 & 0 & A(1+\mu)/L & 0 & 0 \\ 0 & -12I/L^3 & -6I/L^2 & 0 & 12I/L^3 & -6I/L^2 \\ 0 & 6I/L^2 & 2I(1-\mu/2)/L & 0 & -6I/L^2 & 4I(1+\mu/4)/L \end{bmatrix}, \qquad (2)$$

$$\mu = \frac{12EI}{L^2 GA K_s}, \tag{3}$$

$$Q = \begin{bmatrix} n_{x\bar{x}} & n_{y\bar{x}} & 0 & 0 & 0 & 0 \\ n_{x\bar{y}} & n_{y\bar{y}} & 0 & 0 & 0 & 0 \\ 0 & 0 & 1 & 0 & 0 & 0 \\ 0 & 0 & 0 & n_{x\bar{x}} & n_{y\bar{x}} & 0 \\ 0 & 0 & 0 & n_{x\bar{y}} & n_{xy} & 0 \\ 0 & 0 & 0 & 0 & 0 & 1 \end{bmatrix}, \tag{4}$$

where $\bar{K}^e$ is the local element stiffness matrix, and $E, A, I,$ and $L$ are the elastic modulus, the cross-section area, the moment of inertia ($I = Tt^3 / 12$), and the length of the element, respectively. $\mu$ is a dimensionless coefficient that characterizes the importance of shear-related parameters including $G$ (shear modulus) and $K_s$ (shear correction factor = 0.85). $Q$ is the transformation matrix and contains the direction cosines:

$$n_{x\bar{x}} = n_{y\bar{y}} = \frac{x_2 - x_1}{L}, n_{y\bar{x}} = -n_{x\bar{y}} = \frac{y_2 - y_1}{L} \tag{5}$$

where $x_1, y_1, x_2,$ and $y_2$ are the element nodal coordinates.

The element load vector $f^e$ is obtained as follows [30]:

$$f^e = Q^T \bar{f}^e_l, \tag{6}$$

$$\bar{f}^e_l = \begin{bmatrix} q_{\bar{x}} L/2 \\ q_{\bar{y}} L/2 \\ q_{\bar{y}} L^2/12 \\ q_{\bar{x}} L/2 \\ q_{\bar{y}} L/2 \\ -q_{\bar{y}} L^2/12 \end{bmatrix} \tag{7}$$

The stiffness matrix and load vectors of all the elements were calculated and were assembled into a global stiffness matrix ($K$) and a global load vector ($F$). Finally, all the forces and displacements were calculated using Hook's law ($F = Kd$).

To calculate the Young's moduli of the structure ($E_{11} = \frac{\sigma_{11}}{\varepsilon_{11}}$ and $E_{22} = \frac{\sigma_{22}}{\varepsilon_{22}}$), the normal stresses in the directions 1 and 2 ($\sigma_{11} = \frac{\bar{F}_1}{A_2}, \sigma_{22} = \frac{\bar{F}_2}{A_1}$, where $A_1$ and $A_2$ are the cross-section areas of the structure on the 1-3 and 2-3 planes (Fig. 1a)) were divided by the strain applied along the same direction ($\varepsilon_{11} = \varepsilon_{22} = 3\%$). In these equations, $\bar{F}_1$ and $\bar{F}_2$ are, respectively, the mean reaction forces along the directions 1 and 2 at the right and top nodes ($\overline{F_1} = \frac{\sum_{i=1}^{n_R} F_{1_i}}{n_R}, \overline{F_2} = \frac{\sum_{i=1}^{n_T} F_{2_i}}{n_T}$, where $n_R$ and $n_T$ are the total numbers of the right and top nodes while $F_{1_i}$ and $F_{2_i}$ are the reaction forces along the directions 1 and 2 at each of the right and top nodes, respectively). To calculate the Poisson's ratio ($v_{12} = v_{21} = -\frac{\varepsilon_{trans}}{\varepsilon_{axial}}$), the transverse strain was first calculated as the ratio of the mean displacement of the lateral nodes to the initial transversal length of the structure. The transverse strain was then divided by the applied axial strain (in the case of $\varepsilon_{axial} = \varepsilon_{11} = 3\%$: $\varepsilon_{trans} = \varepsilon_{22} = \frac{\sum_{i=1}^{n_T} \delta y_i}{L_2 n_T}$, and in the case of $\varepsilon_{axial} = \varepsilon_{22} = 3\%$: $\varepsilon_{trans} = \varepsilon_{11} = \frac{\sum_{i=1}^{n_R} \delta x_i}{L_1 n_R}$ where $L_1$ and $L_2$ are the initial lengths of the structure along the directions 1 and 2).

## 2.2. Deep learning

We implemented two artificial neural networks (ANN) using Keras libraries, namely the 'single unit cell model' and the 'four-tile model'. The single unit cell model predicted the mechanical properties of the lattice structures with three unit cell angles of 60°, 90°, and 120° and a wide range of $\rho_h$ values (*i.e.*, $\rho_h (\%) = 5, 10, 20, 30, 40, 50, 60, 70, 80, 90$, and $95$). To train the single unit cell model, the finite element models were first solved for 16,500,000 lattice structures with random assignments of the hard phase within the structure. The inputs to the single unit cell deep learning model included 150 material parameters indicating whether each strut was hard or soft (1 = hard, 0 = soft) and one unit cell angle ($\theta$ = 60, 90, and 120) (151 inputs in total). The outputs of the model included the elastic moduli ($E_{11}$, $E_{22}$) and Poisson's ratios ($v_1$, $v_2$) in both directions (4 outputs in total) (Fig. 1a).

The dataset generated for the single unit cell model was employed in the training of the four-tile deep learning model. Towards this aim, 90 mechanical properties were selected from within the achievable range of elastic properties predicted by the single unit cell deep learning model.

All possible four-combinations of these mechanical properties (*i.e.*, $n_2 = C(90, 4) = 2555190$) were generated. Our finite element code was then used to calculate the overall elastic properties of the structures resulting from the four-tile arrangements of these designs (Fig. 1b). The four-tile deep learning model was then created to map the space of the 16 input parameters (*i.e.*, the elastic properties of the individual tiles) to the space of 4 output parameters (*i.e.*, the elastic properties of the four-tile structures).

We scaled all the outputs of the single unit cell models and all the inputs and outputs of the four-tile model to the range [0-1] (see Table 1 for the scaling method). In post-processing, we scaled the relevant outputs back to the original range to facilitate the interpretation of the results. The deep learning models consisted of a series of fully connected layers. A fully connected layer is a function from *m* nodes to *n* nodes. Each output dimension depends on each input dimension. We used the sequential model based on the tf.keras library. To optimize the model, we tuned the values of the hyperparameters (*i.e.*, the number of hidden layers, the number of neurons in each hidden layer, the activation function, the number of the training epochs, and the size of training batches). We randomly selected 98% of the data for the training and the rest for testing the deep learning models. The training dataset was further split into a training sub-set (80%) and a validation sub-set (20%). The validation sub-set was used to tune the hyperparameters of the model. We implemented an early stopping callback from tf.keras library to stop the training process for Four-tile model when there was no improvement in the loss function. The number of the hidden layers, the type of the activation functions, the type of feature scaling and optimization algorithms, the learning rate, and the number of training epochs are presented in Table 1.

In order to calculate the training error, the prediction results of the deep learning models were compared with the target values (finite element simulation results) and the mean absolute error (MAE) as well as the mean squared error (MSE) were calculated for each training epoch. MAE quantifies the magnitude of the prediction error without considering the error direction:

$$MAE = \frac{1}{n}\sum_{i=1}^{n}|y_i - \hat{y}_i| \tag{8}$$

where $n$ is the number of the training samples, $y_i$ are the predicted values, and $\hat{y}_i$ are the true values. The mean squared error is the squared mean of the differences between the predicated values, $y_i$, and the true values, $\hat{y}_i$, and is calculated as:

$$MSE = \frac{1}{n}\sum_{i=1}^{n}(y_i - \hat{y}_i)^2 \tag{9}$$

### 2.3. Experiments

To validate the results of our computational models used for training the single unit cell models, we selected three single unit cell lattice structures (one from each of the cell angles of 60°, 90°, and 120°) (Fig. 1a). In addition, we designed three nine-tile structures and four four-tile structures. These structures represented different arrangements of the single unit cell designs (see Section 3.6). The selected designs were 3D printed and mechanically tested.

We used a multi-material 3D printer (Object500 Connex3, Stratasys, US) which uses the jetting of multiple UV-curable polymers (Polyjet technology) for printing multi-material structures. The commercially available polymers VeroCyan™ (hard phase, RGD841) and Agilus30™ white (soft phase, FLX985) were employed (both from Stratasys, USA). The hard and soft phases were selected such that the ratio of the elastic modulus of the hard phase ($E_h \cong 60$ MPa) to that of the soft phase ($E_s \cong 0.60$ MPa) was around 100. We designed a pin and gripper system to attach the printed specimens to the mechanical testing machine. These parts were 3D printed using a fused deposition modeling (FDM) 3D printer (Ultimaker 2$^+$, Geldermalsen, the Netherlands) from polylactic acid (PLA) filaments (MakerPoint PLA, 750 gr, Natural). A mechanical testing machine (LLOYD instrument LR5K, load cell = 100 N) was used to load the specimens under tension (stroke rate = 1 mm/min). The applied displacement and the reaction force were recorded to obtain the stress-strain curve by dividing the force by the initial cross-section area and dividing the displacement by the initial length of the specimen. The slope of the stress-strain curve represents the overall stiffness of the sample. This procedure was repeated for a total of ten specimens.

We also used the digital image correlation (DIC) technique to measure the full-field strain distribution during the uniaxial tensile tests for the selected single unit cell lattice structures. The surface of the specimens was first painted white. A spackle pattern was then applied to the surface using an airbrush. We used a digital image correlation system (Q400-3D-12MP, LIMESS Messtechnik u. Software GmbH, Germany) equipped with two cameras (DCM 12.0 Mpixel, digital monochrome high performance GigE camera) to record a series of image pairs from two different angles that were later analyzed with the help of the associated commercial software (Istra4D, Germany) to establish the correlations in the images and calculate the full-field strain maps (Fig. 1a).

**3. RESULTS AND DISCUSSION**

**3.1. Training and performance of the deep learning models**

Using a Workstation (CPU = Intel® Core™ i9-8950HK, RAM = 32.0 GB) and one running script, each finite element simulation could be performed between $6.2 \times 10^{-2} \pm 2.7 \times 10^{-3}$ s and $6.5 \times 10^{-3} \pm 8.2 \times 10^{-4}$ s while each deep learning prediction took between $8.3 \times 10^{-2} \pm 2.9 \times 10^{-3}$ s and $1.2 \times 10^{-5} \pm 1.2 \times 10^{-6}$ s depending on the number of simultaneously run simulations/predictions (a comparison between the FE simulation time and the deep learning prediction time for the single unit cell model is presented in Fig. S1 of the supplementary document). The solution time per design also depends on the number of scripts run in parallel. For instance, for $10^5$ simultaneously run simulations and $10^6$ simultaneously run deep learning predictions, each finite element simulation could be performed within $5.0 \times 10^{-3} \pm 4.9 \times 10^{-4}$ s - $2.5\times10^{-3} \pm 1.3\times10^{-4}$ s while each deep learning prediction took between $1.3\times10^{-5} \pm 1.8\times10^{-7}$ s and $2.4\times10^{-6} \pm 1.2\times10^{-7}$ s depending on the number of simultaneously run scripts.

Within 200 epochs of training, the prediction errors (MSE and MAE) of the single unit cell models reduced from $6.6\times10^{-4}$ and $1.38\times10^{-2}$ to $1.05\times10^{-4}$ and $6\times10^{-3}$, respectively. Meanwhile, the prediction errors of the validation dataset decreased from $4.25\times10^{-4}$ and $1.19\times10^{-2}$ to $1.14\times10^{-4}$ and $6.38\times10^{-3}$, respectively. In the case of the four-tile model, the prediction errors (MSE and MAE) corresponding to the training and validation datasets reduced within 27 epochs from (MSE=$1.2\times10^{-3}$, MAE=$2.04\times10^{-2}$) and (MSE=$6\times10^{-4}$, MAE=$1.84\times10^{-2}$) to (MSE=$6.79\times10^{-5}$, MAE=$6.2\times10^{-3}$) and (MSE=$1.1\times10^{-4}$, MAE=$8.1\times10^{-3}$). The coefficients of determination of the single unit cell and four-tile deep learning models were respectively $9.98\times10^{-1}$ and $9.94\times10^{-1}$ (more information is provided in Table S2 of the supplementary document), indicating that these models were highly accurate in predicting the mechanical properties of both types of soft-hard lattices. Given this high degree of accuracy, the deep learning models were used in the rest of the study for evaluating the mechanical properties of the designed structures.

### 3.2. Single unit cell deep learning model

We used the trained 'single unit cell' deep learning model to predict the mechanical properties of $1.5\times10^{9}$ random structures (Fig. 1a). The predicted ranges of the elastic moduli (*i.e.*, $E_{11} \in$ 0 to 10.67 MPa and $E_{22} \in$ 0 to 0.54 MPa) and Poisson's ratios (*i.e.*, $\nu_{12} \in -1.24$ to 1.16 and $\nu_{21} \in -0.53$ to 0.51) were quite broad (more information is provided in Table S3 of the supplementary document). Along direction 1, a wide range of elastic properties (*i.e.*, $E_{11}$, $\nu_{12}$ duos) were obtained within a conifer cone-like region. In comparison, the range of the elastic properties found for direction 2 (*i.e.*, $E_{22}$, $\nu_{21}$ duos) was narrower and included several bean-like regions (Fig. 1a). High elastic modulus ($E_{11}$) values were achieved when orthogonal unit cells were used, which is expected, given that the deformation of orthogonal unit cells under orthogonal loading is primarily stretch-dominated. Highly negative and highly positive Poison's ratios were predicted for the lattices based on the re-entrant and honeycomb unit cells, respectively. $E_{11}$ and the absolute value of $v_{12}$ were inversely correlated for $\rho_h$ values up to 80%, after which they were directly correlated (Fig. 1a). According to the predictions of the Hashin–Shtrikman theory and the theoretical limits established for composite materials [31], [32], an inverse relationship between the elastic modulus and Poisson's ratio is expected. However, the direct correlation observed for the $\rho_h$ values exceeding 80% is caused by the non-affinity imposed by the random distribution of the hard phase within the lattice structures. In another study [33], we showed that the Poisson's ratio and the degree of non-affinity ($\Gamma$) are related to each other through a power law for both re-entrant and honeycomb unit cells. Furthermore, it was concluded that regardless of the type of the unit cell and the level of the applied strain, the degree of non-affinity increases with $\rho_h$ until a maximum value is reached at $\rho_h$ = 75% − 90% after which it decreases to reach $\Gamma = 0$ for the structures only made from the hard phase (*i.e.,* $\rho_h = 100\%$). These statements clearly explain the asymmetry in the plot of $E_{11}$ *vs.* $v_{12}$ for both re-entrant and honeycomb types of the unit cells.

**3.3. Four-tile deep learning model**

Using the four-tile deep learning model, we studied the elastic properties resulting from the various combinations of four tiles with different mechanical properties (Fig. 1b). Along direction 1, the region representing the attainable properties is less symmetric (Fig. 1b) but is nevertheless more so than the one achieved with a single unit cell model (Fig. 1a). Along direction 2, the four-tile model predicted a near-square region representing the attainable elastic properties, which is a remarkable achievement and means that high values of elastic modulus can be combined with highly negative or highly positive values of the Poisson's ratio (Fig. 1b). These observations confirm that a simple four-tile arrangement of the random multi-material designs can greatly expand the achievable range of anisotropic elastic properties.

### 3.4. Role of multi-material design

In the single unit cell model, the ranges of both elastic moduli ($E_{11}$, $E_{22}$) monotonically increased with $\rho_h$ regardless of the type of unit cell (Fig. 2a). This is expected, given that increasing the volume ratio of the hard phase to the soft phase simply increases the elastic modulus of the composite lattice structure. The plots of the Poisson's ratios *vs.* $\rho_h$ were not monotonic with the absolute values of $\nu$ initially increasing until a global extremum was reached for $\rho_h > 50\%$ (*i.e.*, 60-80%), followed by a decreasing trend. For all the three types of unit cells, the ranges of the attainable Poisson's ratios were the widest for $\rho_h = 60 - 80\%$. This is the range where the multi-material nature of the designs plays the most important role in determining the Poisson's ratio of the lattice structure, given that both phases have comparable effects. For smaller or larger values of $\rho_h$, either the soft or the hard phase dominates the mechanical response of the lattice structure, respectively. For a fixed value of the Poisson's ratio (*i.e.*, $\nu_{12} = -1 \pm 0.01, \nu_{12} = 0 \pm 0.01$, and $\nu_{12} = 1 \pm 0.01$), a wide range of elastic moduli were achieved, depending on the type of unit cell and $\rho_h$ (Fig. 2b). For fixed values of $\nu$ and $\rho_h$, the largest range of the elastic moduli was achieved for the larger $\rho_h$. For example, for the designs with orthogonal unit cells and with a $\rho_h$ value of 80%, the elastic modulus can change by up to 10.7 folds, depending on how the hard and soft phases are assigned to the lattice structure and without any noticeable change in the Poisson's ratio (*i.e.,* $\nu_{12} = 0 \pm 0.01$). This highlights the importance of multi-material design aspect in the tunability of the elastic properties of mechanical metamaterials.

### 3.5. Stiff double-auxetic structures

We also studied how the assignment of hard and soft phases in multi-material lattices as well as combining different types of unit cells in a four-tile structure could be used to achieve double-auxetic, yet stiff structures. When combining different types of unit cells, one of the chosen unit cell types should always be the re-entrant unit cell, leading to four possible combinations. The predictions of our deep learning models indicate that combining these unit cells enables us to achieve double-auxetic, yet stiff lattice structures (Fig. 2c). Double-auxeticity is a rare event on its own [18], let alone combined with high stiffness, further underscoring the importance of the implemented design strategies. Furthermore, the presented combinations of different unit cell types enable the coverage of a substantial part of the $(E_{11}, E_{22})$ and $(\nu_{12}, \nu_{21})$ planes.

**3.6. Tiled and transformed structures**

We selected the following single unit cells designs for a more in-depth study: a design with the highest values of $E_{11}$ and $E_{22}$ from the orthogonal unit cell group ($E_{11} = 9.67$ MPa, $E_{22} = 0.31$ MPa, $\nu_{12} = -0.04$, and $\nu_{21} = 0.00$), a design with the most negative value of the Poisson's ratio and almost the highest elastic modulus from the re-entrant unit cell group ($E_{11} = 0.93$ MPa, $E_{22} = 0.26$ MPa, $\nu_{12} = -1.17$, and $\nu_{21} = -0.45$), and a design with the most positive value of the Poisson's ratio and an almost highest elastic modulus from the honeycomb unit cell group ($E_{11} = 1.31$ MPa, $E_{22} = 0.52$ MPa, $\nu_{12} = 1.05$, and $\nu_{21} = 0.47$). We then arranged these designs into four-tile (Fig. 3a) and nine-tile (Fig. 3b) structures and obtained their mechanical properties and deformation patterns both computationally and experimentally. Moreover, we studied how a 90-degree rotation of a design would affect the mechanical properties of the combined structures (Fig. 3a). We found that combining the abovementioned designs further expanded the space of achievable elastic properties, filling the gaps in mechanical properties of individual unit cells. For instance, in structure 1 (Fig. 3a) and structure 5 (Fig. 3b), the combination of re-entrant and orthogonal unit cells boosted the elastic modulus ($E_{11}$) of the constituent re-entrant unit cell by 75.6% and 91.4%, respectively, while the Poisson's ratio maintained its extreme negative values ($|\nu_{12}|$ reduced by 6.6% and 9.4%, respectively). In structure 2 (Fig. 3a) and structure 6 (Fig. 3b), the combination of honeycomb and orthogonal unit cells boosted the elastic modulus ($E_{11}$) of the constituent honeycomb unit cell by 40% and 40.4%, respectively, while the extreme positive Poisson's ratios did not change much ($|\nu_{12}|$ reduced only by 1.4% and 2.8%, respectively).

We also showed that with a 90-degree rotation of a design, we could increase the elastic modulus in the weak direction ($E_{22}$) and create structures with a higher level of isotropy. In this way, we could achieve structures with a higher elastic modulus ($E_{22}$) than both types of their constituent designs. For example, the elastic modulus ($E_{22}$) of structure 3 was 56.2% and 31% higher than the elastic modulus ($E_{22}$) of the constituent re-entrant and orthogonal unit cells, respectively (Fig. 3a). In structure 4 (Fig. 3a), the elastic modulus ($E_{22}$) was 4.5% and 75.3% higher than the elastic modulus ($E_{22}$) of the constituent honeycomb and orthogonal unit cells, respectively.

We also showed how the change of boundary conditions would affect the deformation patterns and also contributed to a more uniform stress distribution within the lattice structure. In all designs, the experimental observations regarding the deformation patterns as well as the experimental values of the mechanical properties clearly agreed with our computational results (Table 2), confirming the validity of the computational approach used here.

Furthermore, the combination of structures with different types of unit cells allows for different functionalities. For instance, the hybrid combination of negative Poisson's ratios with positive values could be used to design orthopedic implants with improved longevity [34]. Combining different types of unit cells could create action-at-a-distance behavior that enables different patterns of local actuation using a single far-field deformation and has various potential applications in soft robotics [35]. Here, we also showed that combining different unit cells allows for shape-morphing boundaries as well as for specific values of the Poisson's ratio. For instance, different shape-morphing boundaries were observed in structure 7 (Fig. 3b) when re-entrant and honeycomb unit cells were combined with each other, while the designed structure had a zero value of the Poisson's ratio in both directions. Such properties are of high interest in high added value industries, such as the biomedical and aeronautical industries, as they exhibit improved damping performance [36].

### 3.7. Uniformity of stress distribution

To date, most studies on mechanical metamaterials have focused on the elastic properties of architected lattices without paying much attention to the structural integrity aspects including the risk of failure due to such phenomena as stress concentrations. Generally speaking, the presence of stress concentration leads to premature failure caused by premature initiation and growth of cracks. It is, therefore, desirable to distribute the stresses as uniformly as possible within the lattice structure. An important advantage of having giga-sized databases of possible designs with the corresponding elastic properties is the possibility to apply additional design criteria, such as the one related to the uniformity of the stress distribution.

For example, among all the designs with the same range of predicted elastic properties (*i.e.*, $0.15 < E_{11} < 0.25, -1.1 < \nu_{12} < -1, 0.01 < E_{22} < 0.08,$ and $-0.45 < \nu_{21} < -0.3$), we studied the uniformity of the stress distributions within the lattice structure. In total, 207 tiles with various $\rho_h$ values (*i.e.*, $50, 60, 70, and\ 80\%$) were included (Fig. 4a). The maximum values of the von Mises stress in the structural elements of these designs were calculated while these designs were subjected to two different boundary conditions (*i.e.*, $\varepsilon_{11} = 3\%$ or $\varepsilon_{22} = 3\%$) (Fig. 4b). Although the elastic properties of these designs were generally very similar, the maximum von Mises stresses in their struts varied up to 2.5 and 6.5 times along the loading conditions 1 and 2, respectively. This finding indicates the importance of applying an additional design rule regarding the stress uniformity within the structure. For that reason, two designs with $\rho_h = 70\%$ (one with the minimum and one with the maximum Euclidean distance from the origin) were selected for a more in-depth analysis (Fig. 4b and 4c). A closer study of the stress distributions in these two structures showed a clear incident of stress concentration in design 2 while design 1 exhibited more uniform stress distributions (Fig. 4d). Such types of stress risers are the primary zones for crack initiation and will ultimately result in premature fracture. It is, therefore, important to consider stress uniformity as an additional design requirement in the design of mechanical metamaterials. It should also be mentioned that the maximum von Mises stresses in soft and hard struts of these selected designs are lower than the tensile strengths of the individual materials in the bulk form.

We also studied the distribution of the compressive or tensile axial stresses ($S_{11}$) in individual struts of the selected designs under the aforementioned boundary conditions. That included $S_{11@\varepsilon_{22}=3\%}$ vs. $S_{11@\varepsilon_{11}=3\%}$ values as well as 95% confidence ellipses fitted to the stress values of individual struts for the multi-material designs (Fig. 4e). We then compared these results with the axial stresses obtained from a lattice structure with $\rho_h = 70\%$ and equivalent homogenous material properties (Fig. 4e). This comparison highlighted that the lattice design with a lower stress riser point (*i.e.,* design 1) was located inside the confidence ellipse of the design with the equivalent homogenous material (Fig. 4e). Such an approach can, therefore, be considered as additional design rule for selecting the optimum multi-material design with target properties.

## 4. CONCLUSIONS

In conclusion, deep learning models can accurately predict the mechanical properties of multi-materials mechanical metamaterials, reduce the speed of evaluating each design, and make parallel computing efficient and straightforward to the point where evaluating $10^{10}$-$10^{20}$ designs is within reach. Our results show that such unprecedented sizes of the design database enable the rational design of multi-material mechanical metamaterials that not only achieve a very wide range of elastic properties but also meet additional design requirements. For example, we demonstrated that double-auxetic yet stiff designs can be realized using this approach. Another application demonstrated here is the addition of a criterion regarding stress uniformity that can reduce stress concentration in such types of mechanical metamaterials, thereby increasing their fracture and fatigue resistance.

COMPETING INTERESTS

The authors declare no competing interests.

ACKNOWLEDGEMENTS

The work is part of the 3DMED project that has received funding from the Interreg 2 Seas program 2014–2020, co-funded by the European Regional Development Fund under subsidy contract No. 2S04-014.

**Table 1.** The training parameters of the single unit cell and four-tile deep learning models.

| Parameters | Single-tile model | Four-tile model |
|---|---|---|
| Hidden layer dimensions | 256-128-128-64-32 | 256-128-64-32-16-8 |
| Activation function | ReLU | ReLU |
| Feature scaling | min-max | min-max |
| Optimization algorithm | RMSprop | RMSprop |
| Learning rate | $10^{-4}$ | $10^{-4}$ |
| Number of epochs | 200 | 27 |

**Table 2.** Comparisons between the computationally determined and experimentally measured elastic properties of the multi-tile designs.

| Type | Structure number | FE simulation | | | | Experimental test | | | |
|---|---|---|---|---|---|---|---|---|---|
| | | $E_{11}$ [MPa] | $E_{22}$ [MPa] | $v_{12}$ | $v_{21}$ | $E_{11}$ [MPa] | $E_{22}$ [MPa] | $v_{12}$ | $v_{21}$ |
| Four-tile | 1 | 1.63 | 0.22 | -1.10 | -0.21 | 1.64 | 0.38 | -0.96 | -0.32 |
| | 2 | 1.83 | 0.28 | 1.03 | 0.22 | 1.77 | 0.19 | 0.95 | 0.27 |
| | 3 | 1.02 | 0.41 | -0.74 | -0.31 | 0.99 | 0.57 | -0.64 | -0.25 |
| | 4 | 1.17 | 0.54 | 0.76 | 0.38 | 1.08 | 0.35 | 0.68 | 0.26 |
| Nine-tile | 5 | 1.78 | 0.21 | -1.06 | -0.17 | 1.13 | 0.26 | -0.91 | -0.21 |
| | 6 | 1.84 | 0.26 | 1.08 | 0.21 | 1.78 | 0.24 | 0.96 | 0.15 |
| | 7 | 0.83 | 0.24 | 0.29 | 0.08 | 0.77 | 0.27 | 0.26 | 0.11 |

**FIGURE CAPTIONS**

**Figure 1.** The structures of two optimized deep learning models, namely single unit cell model (a) and four-tile model (b), for designing multi-material lattice structures as well as the relevant training procedures and prediction results. The strain distribution and deformation patterns obtained from FEM and DIC for selected representative designs are presented in the subfigure (a). The prediction *vs.* simulation results and the coefficients of determination for the test data sets are respectively presented in Fig. S2a-b of the supplementary document for the single unit cell and four-tile models.

**Figure 2.** (a) The elastic properties ($E_{11}, \nu_{12}, E_{22}$, and $\nu_{21}$) of the different types of unit cells as predicted by the single-tile deep learning model for the different values of $\rho_h$ (b) The achievable range of the elastic moduli for some specific values of the Poisson's ratio (*i.e.*, $\nu_{12} = -1 \pm 0.01, \nu_{12} = 0 \pm 0.01$, and $\nu_{12} = 1 \pm 0.01$) considering the different values of $\rho_h$ (c) The elastic properties achieved by the single unit cell and four-tile models with a focus on double-auxetic designs. The magnified view shows the distribution of double-auxetic structures by types of unit cells of constituent designs.

**Figure 3.** Different combinations of designs with extreme mechanical properties were selected from each group of the unit cells (*i.e.*, a re-entrant structure with a highly negative Poisson's ratio, a honeycomb structure with a highly positive Poisson's ratio, and an orthogonal structure with a high value of the elastic modulus). Subfigure (a) shows the mechanical properties of two four-tile structures with non-rotated and rotated tiles and the distribution of the von Mises stresses in these lattice structures. These multi-material 3D printed specimens were mechanically tested in both the 1- and 2-directions under 3% tensile strain and the experimental results were compared with the finite element simulation results (Table 2). Subfigure (b) shows the mechanical properties of nine-tile combinations and the von Mises stress distribution in these combinations. These multi-material 3D printed specimens were mechanically tested in both the 1- and 2-directions under 3% tensile strain, and the experimental results were compared with the finite element simulation results.

**Figure 4.** Selecting the design with the most uniform stress distribution among all the designs with similar elastic properties. Subfigure (a) shows the selection of a certain range of the elastic properties within the possible range of the elastic properties predicted by the single unit cell model for the unit cells with a cell angle of 60°. In total, the elastic properties of 207 designs with various $\rho_h$ values (*i.e.*, 50, 60, 70, and 80) fall within these selected ranges. Subfigure (b) shows the maximum values of the von Mises stress in the structural elements of the corresponding lattice structures when these structures were subjected to two boundary conditions (*i.e.*, $\varepsilon_{11} = 3\%$ or $\varepsilon_{22} = 3\%$). Two designs were selected, including one with the minimum (node 1) and one (node 3) with the maximum Euclidean distance from the origin (subfigure (b) and (c)). The distribution of the von Mises stresses in the selected designs and their deformations under two boundary conditions (*i.e.*, $\varepsilon_{11} = 3\%$ and $\varepsilon_{22} = 3\%$ ) are presented in subfigure (d). The axial stresses ($S_{11}$) in the struts of each design under two boundary conditions (*i.e.*, $\varepsilon_{11} = 3\%$ or $\varepsilon_{22} = 3\%$ ) are calculated and are compared with the axial stresses of the corresponding struts when the lattice structure is composed of an equivalent homogenous material (e).

**Figure 2**

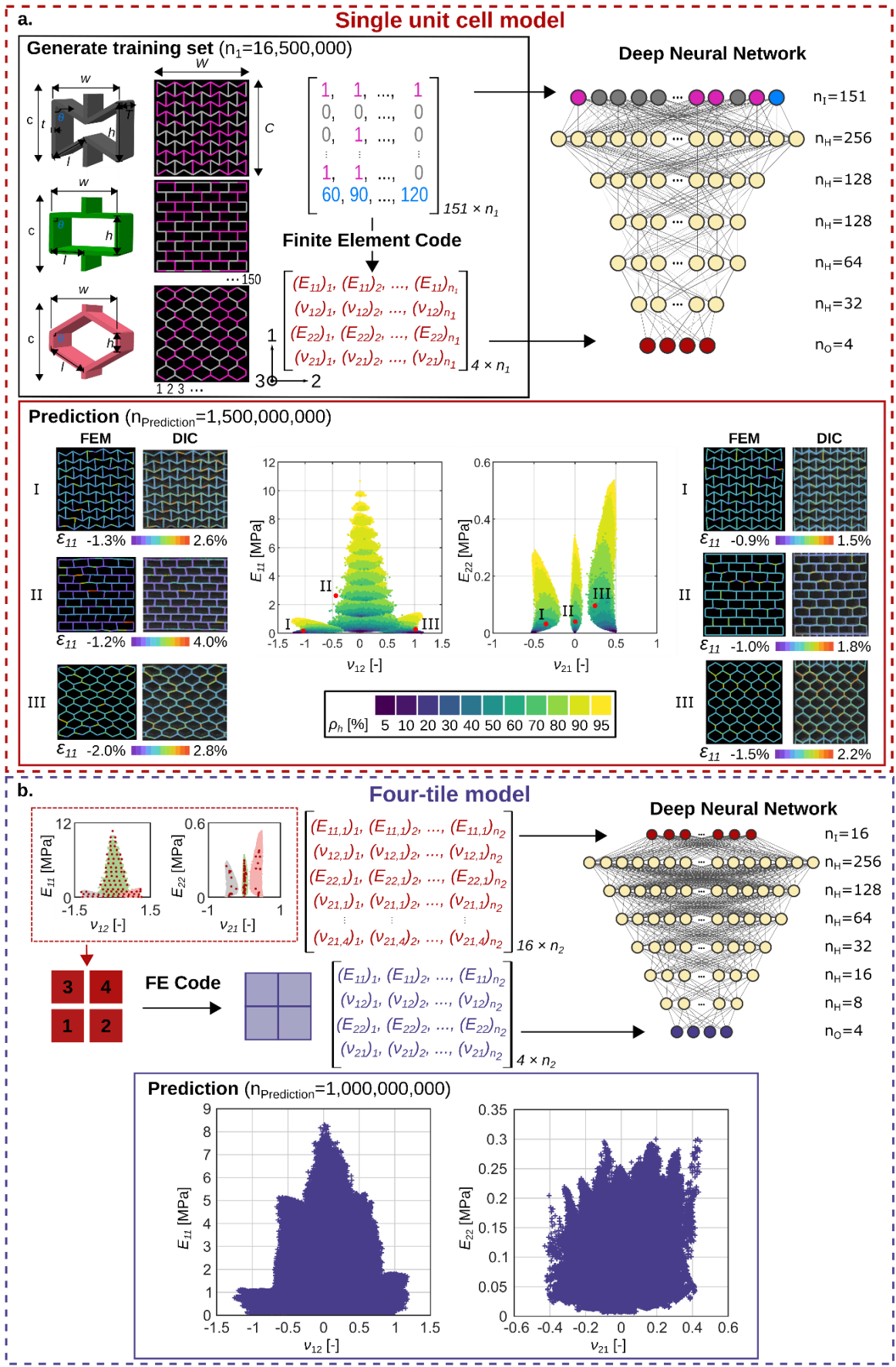

**Figure 2**

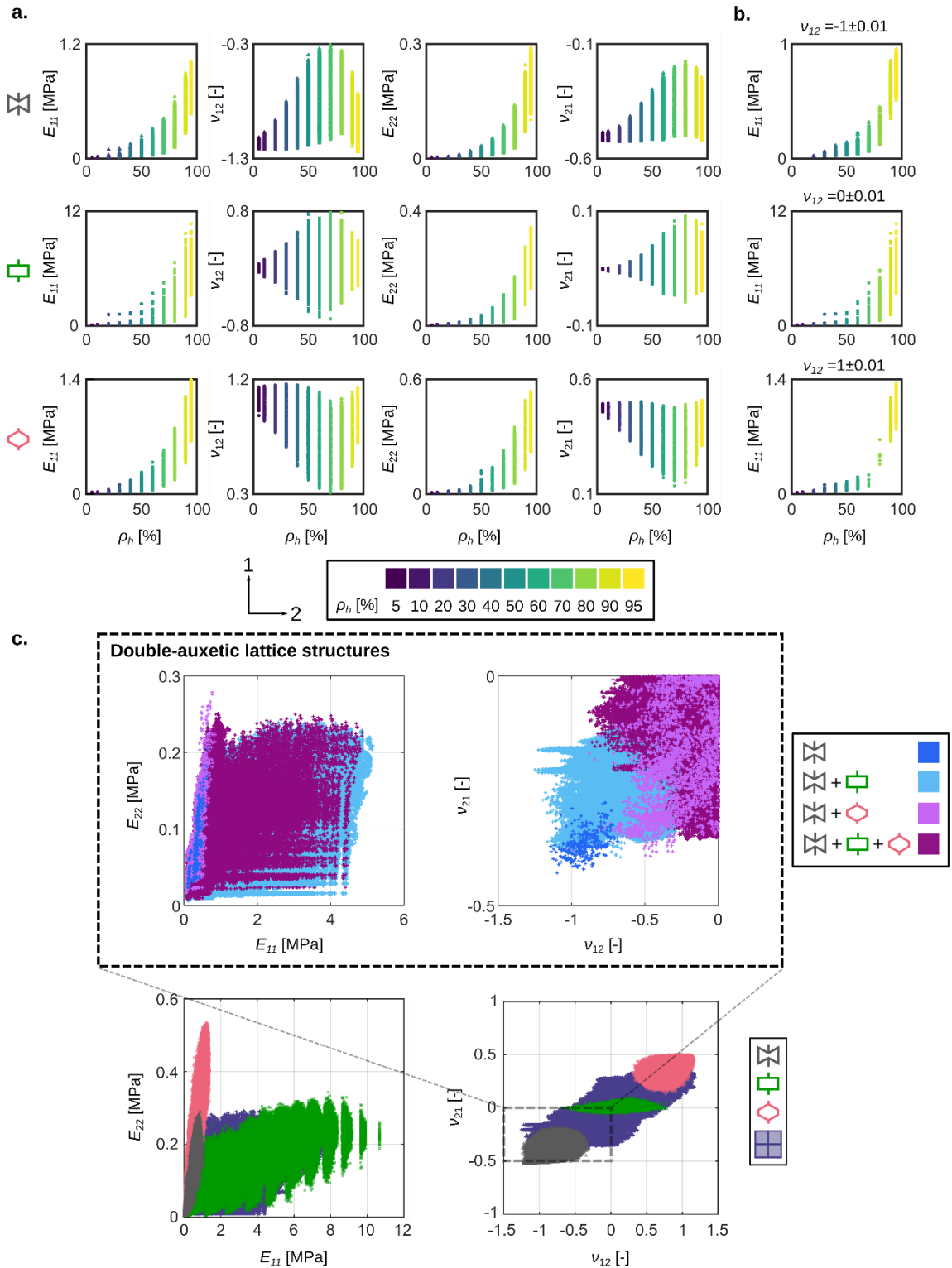

**Figure 3**

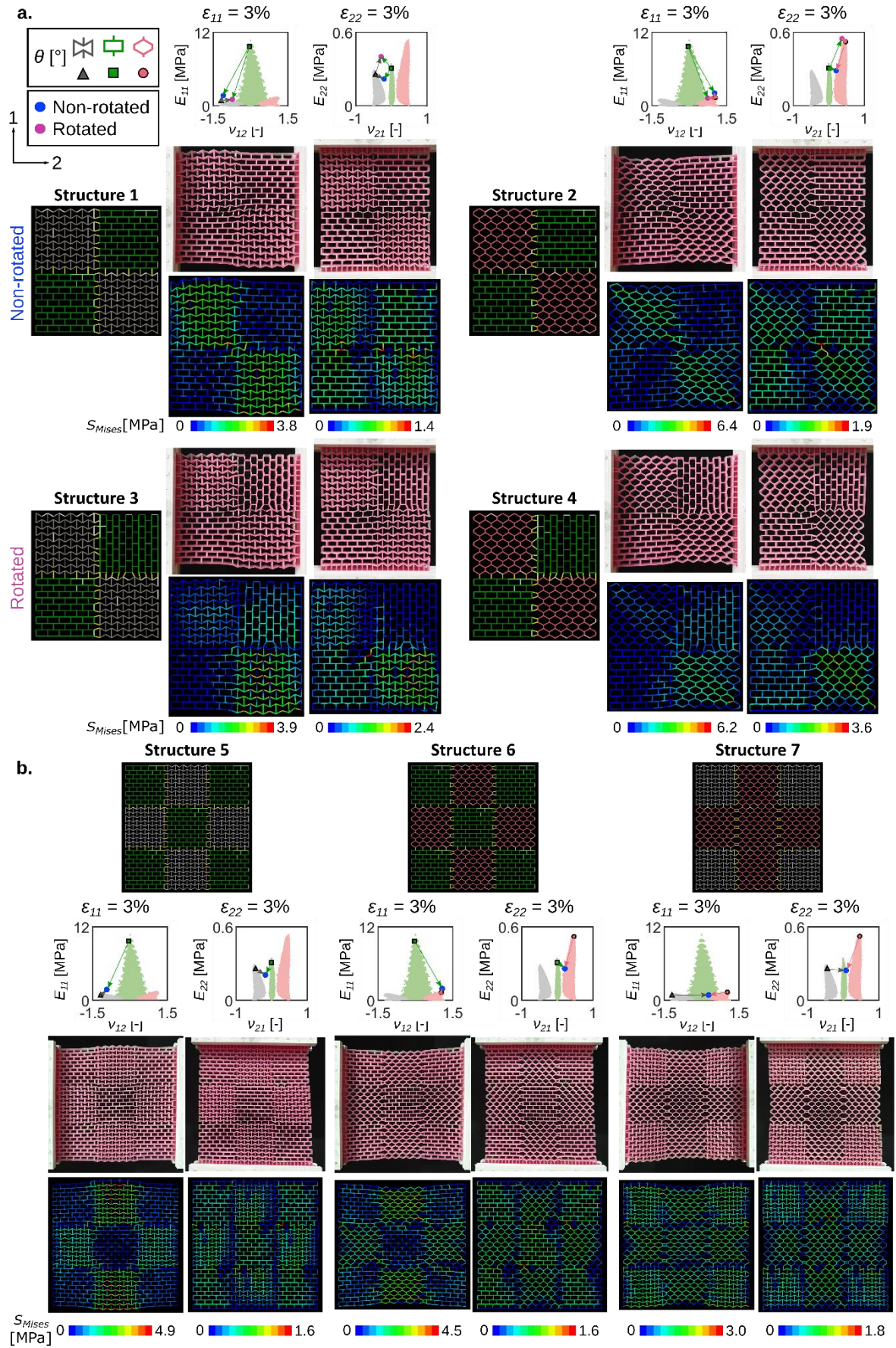

**Figure 4**

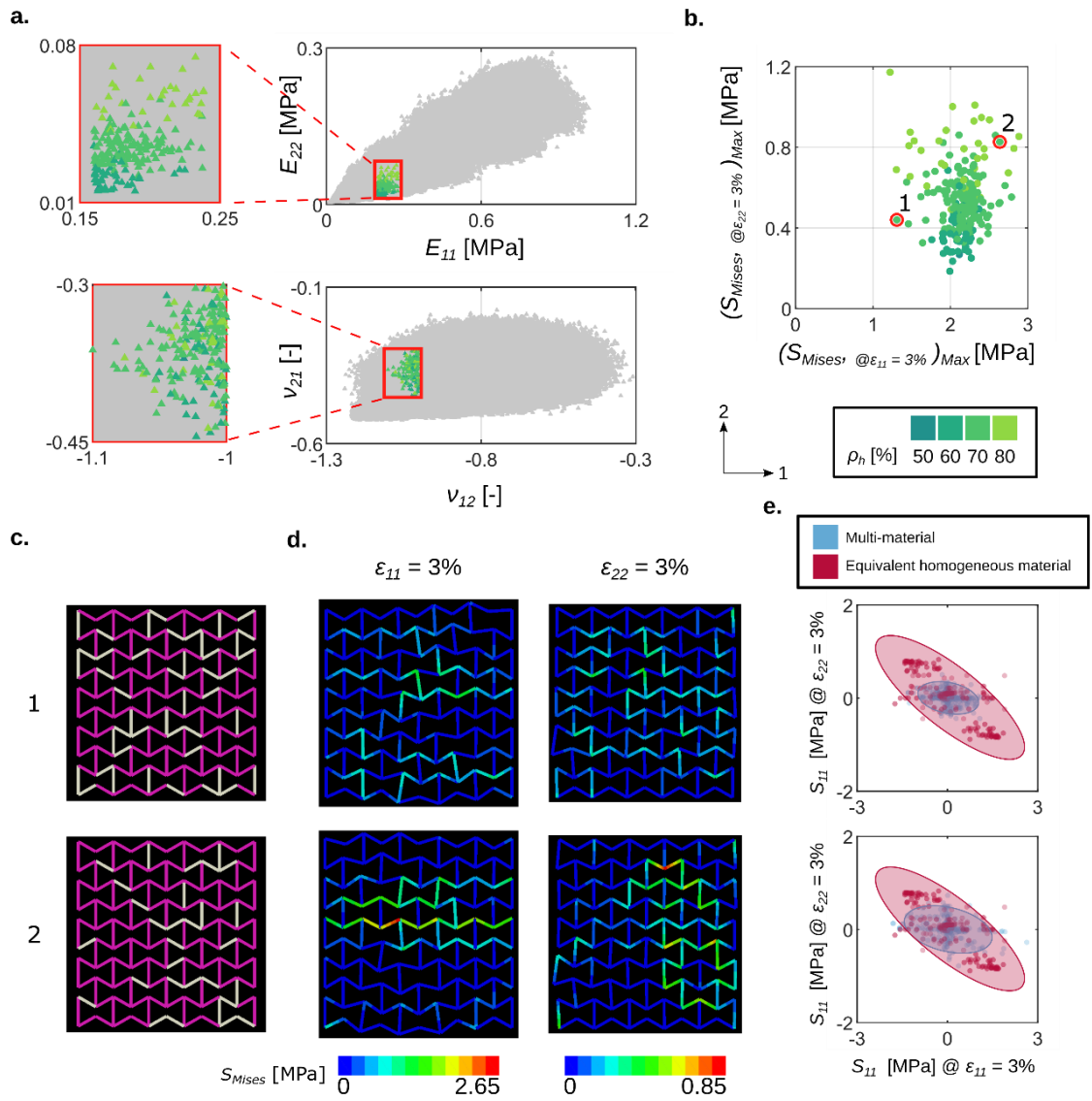